\documentclass[prd,twocolumn,showpacs,preprintnumbers]{revtex4}
\pdfoutput=1
\usepackage{amsfonts,amsmath,amssymb}
\usepackage{graphicx}
\usepackage{color}
\usepackage{natbib}
\usepackage{lipsum}
\usepackage[plainpages=false, colorlinks=true, anchorcolor=blue, linkcolor=blue, citecolor=blue, bookmark
s=false]{hyperref}

\begin{document}

\newcommand{\apjl}{Astrophys. J. Lett.}
\newcommand{\apjs}{Astrophys. J. Suppl. Ser.}
\newcommand{\aap}{Astron. \& Astrophys.}
\newcommand{\aj}{Astron. J.}
\newcommand{\araa}{Ann. Rev. Astron. Astrophys. } 
\newcommand{\mnras}{Mon. Not. R. Astron. Soc.}
\newcommand{\solphys}{Solar Phys.}
\newcommand{\jcap}{JCAP}
\newcommand{\pasj}{PASJ}
\newcommand{\pasa}{Pub. Astro. Soc. Aust.}
\newcommand{\apss}{Astrophysics \& Space Science}
\newcommand{\aaps}{Astron. Astrophys. Suppl. Ser.}
\newcommand{\Dtwoo}{$\mathrm{D_2O}$ }

\title{A search for evidence  of solar rotation in Super-Kamiokande  solar neutrino dataset}
\author{Shantanu  \surname{Desai}$^{1,2}$} \altaffiliation{E-mail: shntn05@gmail.com}
\author{Dawei W.   \surname{Liu}$^{3}$} \altaffiliation{E-mail: liudawei@gmail.com}
\affiliation{$^{1}$Excellence Cluster Universe, Boltzmannstr. 2, D-85748, Garching, Germany}
\affiliation{$^{2}$Ronin Institute, Montclair, NJ, USA }
\affiliation{$^{3}$Department of Physics, University of Houston, Houston, TX, USA }

\begin{abstract}

We apply the generalized Lomb-Scargle (LS) periodogram, proposed by Zechmeister and Kurster, to the solar neutrino data from Super-Kamiokande (Super-K) using data from  its first five years. For each peak in the LS periodogram, we evaluate the statistical significance in two different
 ways. The first method involves calculating the False Alarm Probability (FAP) using non-parametric bootstrap resampling, and the second method is by calculating the difference in Bayesian Information Criterion (BIC) between the null hypothesis, viz.  the data contains only noise, compared to the hypothesis that  the data contains a peak at a given frequency. Using these methods, we scan the frequency range between 7-14 cycles  per year  to  look for any peaks caused by solar rotation, since this is the proposed explanation for the statistically significant peaks found by Sturrock and collaborators in the Super-K dataset.  From our analysis, we do confirm that similar to Sturrock et al, the maximum peak occurs at a   frequency of 9.42/year, corresponding to a period of 38.75 days. The FAP for this peak is about 1.5\% and the difference in BIC (between pure white noise and this peak) is about 4.8. We note that the significance depends on the frequency band used to search for peaks and hence it is important to use a search band appropriate for solar rotation. The significance of this peak based on the value of BIC  is  marginal and more data is needed to confirm if the peak persists and is real.

\end{abstract}

\maketitle

\section{Introduction}

Ever since the first observations of solar neutrinos~\cite{Davis}, there have been a number of  claimed detections of periodicities in various solar neutrino datasets, based on works from individual research groups~\cite{Press,Sturrock97,Caldwell03,Milsztajn,Ghosh,Ghosh05,Sturrock03a,Sturrock01,Sturrock03,Sturrock04,Sturrock05a,Sturrock06,Sturrock07,Sturrock09,Sturrock12,Sturrock15}.  A variety of theoretical explanations have been proposed to explain  these results, such as anti-correlations with solar activity~\cite{Press}, resonant spin-flavor precession due to non-zero magnetic moment~\cite{Akhmedov}, $r$-mode oscillations induced by solar rotation~\cite{Sturrock97,Sturrock08,Sturrock12}, temperature variation of the solar core~\cite{Raychaudhuri71}, etc. However, these suggestions have not been corroborated by the experimental collaborations. The current solution to the solar neutrino puzzle is the large mixing angle Mikheyev-–Smirnov–-Wolfenstein  neutrino oscillations~\cite{Maltoni}. If these periodicities are confirmed, they could shed insight on new physics  in the neutrino sector, or point to  deficiencies with the Standard Solar Model~\cite{ssm}.

Over the last decade or so, Sturrock and collaborators have argued for the detection of multiple statistically significant peaks in the measured $^8$B neutrino flux by Super-Kamiokande (Super-K). They have shown that the  most significant peaks have amplitude variations of about 7\% and frequencies of  about 9.43/year  and 26.5/year  in the 5-day and 10-day datasets, corresponding to periods of 38.75  and 13.76 days respectively. (See ~\cite{Sturrock01,Sturrock03,Sturrock04,Sturrock05a,Sturrock06,Sturrock07,Sturrock09,Sturrock12,Sturrock15} and references therein).  The most likely explanation of the peak at 9.43/year is due to the synodic rotation  of  the solar core, corresponding to a sidereal rotation rate of   10.43/year in the solar radiative zone~\cite{Sturrock08,Sturrock12}  and the  peak at 26.57/year is an alias of the peak at 9.43/year~\cite{Caldwell03}.
Periodicities correlated with solar rotation rate have also been reported in the Chlorine and Gallium experiments used to detect solar neutrinos such as Homestake, SAGE, GALLEX, and GNO~\cite{Sturrock97,Sturrock01,Sturrock06,Ghosh05,Sturrock09}.

However, these results have not been confirmed by the analysis done by the Super-K 
collaboration~\cite{Yoo}.  The Super-K collaboration looked for periodic modulations from its first five years of solar neutrino data,    after dividing the data into 5-day and 10-day bins using the Lomb-Scargle periodogram and ruled out evidence for modulations at 13.76 and 38.75 days, and also at  all other frequencies~\cite{Yoo}.

Given the controversial and conflicting nature of these claims, it is important for an independent check of the same dataset using different analysis methods and techniques. The analysis carried out by the Super-K collaboration uses the Lomb-Scargle periodogram and calculates the false alarm probability according to the procedure outlined in ~\cite{NR}. Meanwhile, there have been fundamental algorithmic improvements to the original Lomb-Scargle algorithm and also advances in the calculation of the significance of a detected peak, after  the original papers by the experimental collaborations. Also the availability of large scale computing facilities allows these more robust techniques to be easily applied to any datasets. We therefore use modern variants of these techniques to analyze the publicly available Super-K solar neutrino dataset,  in order to confirm or refute these various claims of periodicities. Furthermore, following the cogent arguments made by Sturrock et al for correlations between the peaks in Super-K data and solar rotation, we only look for these 
peaks. The solar rotation rate varies between approximately 13.8/year (at the equator) to  7.7/year (at the poles)~\cite{Beck2000}. Therefore, we restrict our searches  for peaks in the Super-K data in a frequency range between 7-14/year.

The outline of this paper is as follows. In Section I, we briefly review our implementation of the generalized Lomb-Scargle periodogram, including the calculation of significance of any detected peaks.  In Section II, we report  our analysis of the Super-K dataset and present results for the significance of all the peaks found by Sturrock and collaborators on the same datasets. Section III contains our conclusions. In the appendix we look for similar periodicities in the SNO data and also do an estimate of the sensitivity analysis of finding a peak in the Super-K solar neutrino datasets (binned in 5-day and 10-day intervals).

\section{Generalized Lomb-Scargle Periodogram}
The Lomb-Scargle (hereafter, LS)~\cite{Lomb,Scargle} periodogram is a widely used technique in astronomy and particle physics to look for  periodicities in unevenly sampled datasets, and has been applied to a large number of systems from exoplanets to  variable stars.   It can obtained as an analytic solution,  while solving the problem of   fitting for a sinusoidal function by $\chi^2$ minimization, 
 and hence is a special case of the maximum likelihood technique for symmetric errors~\cite{Ranucci05}. Unlike maximum-likelihood  techniques, the original LS periodogram does not account for  the start and end times of each bin or any asymmetric errors, which may exist for any given dataset.

Here, we shall apply the generalized LS periodogram and follow the treatment and notation from {\tt astroML}~\cite{astroml}, where more details can be found. We first introduce the normal LS periodogram and then discuss the modification proposed by Zechmeister and Kurster~\cite{Kurster}, which is known in the literature as the generalized LS periodogram.
We shall also illustrate some of the key differences between this method and the LS periodogram analysis carried out by the Super-K collaboration, which followed the  treatment in ~\cite{NR}. 

The goal of the LS periodogram algorithm is to determine the angular frequency ($\omega$) of a  periodic signal in a time-series dataset  $y(t)$
given by $y(t)=a\cos(\omega t)+ b \sin(\omega t)$. If we consider a  series of measurements \{$y_i$\} at times \{$t_i$\} with uncertainties \{$\sigma_i$\}, the power at a given angular frequency ($\omega$) for this dataset is given by:
\begin{equation}
P(\omega)= \frac{1}{2}\left[\frac{R^2(\omega)}{C(\omega)}+ \frac{I^2(\omega)}{S(\omega)}\right], 
\label{eq:P}
\end{equation}
where 
\begin{eqnarray}
R(\omega) &=& \sum \limits_{i=1}^{N} w_i (y_i-\bar{y}) \cos[w_i(t_i-\tau)] , \\
I(\omega) &=& \sum \limits_{i=1}^{N} w_i (y_i-\bar{y}) \sin[w_i(t_i-\tau)] , \\
C(\omega) &=& \sum \limits_{i=1}^{N} w_i \cos^2 [w_i(t_i-\tau)] , \\
S(\omega) &=& \sum \limits_{i=1}^{N} w_i \sin^2 [w_i(t_i-\tau)] ,   \\
\label{eq:ls}
\end{eqnarray}
and $\bar{y}$ and $w_i$ are given by:
\begin{equation}
\bar{y}  = \sum \limits_{i=1}^{N} w_i y_i, 
\label{eq:bary}
\end{equation}
\begin{equation}
w_i = \frac{1/\sigma_i^2}{1/\sum\limits_{i=1}^N \sigma_i^2}. 
\label{eq:sigma}
\end  {equation}

$\tau$ is calculated via
\begin{equation}
\tan (2\omega \tau) = \frac{\sum \limits_{i=1}^N w_i \sin(2\omega t_i)}{\sum \limits_{i=1}^N w_i \cos(2\omega t_i)} ,
\end{equation}


One assumption in the above formalism of  LS periodogram, is that $\bar{y}$ (computed in Eq.~\ref{eq:bary}) is a good estimate for the mean value of all the observations and is same as that of the fitted function. This may not be correct if the data does not uniformly sample all the phases, or if the dataset is small and does not extend over the full duration of the sample. Such errors in estimating the mean can cause aliasing problems~\cite{Kurster}.
Therefore, to circumvent these issues, the LS periodogram 
was generalized to add an arbitrary offset to the mean values~\cite{Kurster,Cumming}. We refer to this modification  as the ``generalized'' LS periodogram~\cite{Kurster}. We note that similar variants of the generalized LS periodogram involving fitting for a sinusoid followed by an offset have also been proposed in literature under different names,   such as  ``floating-mean periodogram''~\cite{Cumming}, date-compensated discrete Fourier Transform~\cite{Ferraz-Mello}, and ``spectral significance''~\cite{Reegen}. The resulting equations have the same structure  as in Eq. 1-6  and can be found in 
Eq. 20 in  Ref.~\cite{Kurster}. Such a variant has also been used in the  analysis of solar neutrino data~\cite{Sturrock05a,Ranucci05} and is called ``floating-offset'' periodogram in those papers.
It has been shown that the generalized LS periodogram is more sensitive than the normal one in detecting periodicities, in case the data sampling overestimates the mean~\cite{astromlbook,Kurster}. In this work, we shall use the generalized LS periodogram for all analyses.

\subsection{Assessment of Statistical Significance}

For each peak in the LS periodogram, we evaluate the statistical
significance in two different ways. First method  is to use the Bayesian Information Criterion (BIC)~\cite{BIC}, which compares two  models with different number of free parameters. BIC penalizes models with additional complexity or extra free parameters. This is not usually accounted for in traditional maximum likelihood or $\chi^2$ minimization based analyses, where one can always get a better fit with more complex models.
The value of BIC for a given hypothesis is given by:
\begin{equation}
\mathrm{BIC}=\chi^2_{min} + k \ln (N),
\end{equation}

\noindent where $k$ is the number of free parameters, and $N$ is the total number of data points. BIC is widely used in astrophysics and cosmology for model comparison (eg. ~\cite{Liddle,Shi,Shafer} and references therein). However, the absolute value of BIC is only of academic interest and for model comparison, one typically compares the relative values of BIC between two hypotheses, and the one with a smaller value of BIC is considered as the better model. 
Therefore, in this case, to assess the significance of each peak, we compute the difference in BIC between the null hypothesis (assuming that the signal is consistent with a constant mean and white noise) and the value of BIC for a peak detected at  given frequency ($\omega$): 
\begin{equation}
\Delta\mathrm{BIC} = \mathrm{BIC (noise)} - \mathrm{BIC (\omega)}
\label{deltabic}
\end{equation}

For purely homoscedastic errors, $\Delta$BIC reduces to a simple analytic expression related to the  power~\cite{astromlbook}. We shall calculate $\Delta$BIC for each peak in the LS periodogram, after including  the heteroscedastic errors to assess the significance. Therefore, $\Delta$BIC has to be $>0$ for a periodic modulation to be favored  compared to pure noise.
Although, it is difficult to calculate a quantitative $p$-value from $\Delta$BIC, there is a proposed strength of evidence rule to qualitatively assess the relative merits of two models~\cite{Shi} and as a rule of thumb,  $\Delta\mathrm{BIC}>10$ for the hypothesis with a smaller value of BIC to be  statistically significant compared to the one with the larger value. Although, in this work we apply BIC to assess the significance of peaks from our LS analysis, we note that BIC can also be used for model comparison in  generalized likelihood based methods.

The second  method  to calculate the  significance level of a peak at a given frequency ($\omega$) is to calculate  the false alarm probability of the null hypothesis (FAP) for the power at that frequency.  This is   usually  calculated assuming that $P(\omega)$ has an exponential probability distribution with zero mean. The significance is then given by $1-(1-e^{-P(\omega)})^M$~\cite{Scargle,NR}, where $M$ is the number of independent frequencies scanned. For a peak to be considered statistically significant, the FAP has to be  close to 0.
One problem with this approach is that the null hypothesis does not always satisfy the above distribution because of tails in the underlying distribution of the periodogram  and the  number of independent frequencies cannot be precisely defined~\cite{Suveges,Scargle10}. To circumvent these issues, Suveges~\cite{Suveges} has proposed computing the significance (or FAP) using non-parametric bootstrap resampling, which can reproduce any empirical distribution along with extreme-value methods to account for the tails. To assess the significance of any peak, we  shall compute the significance using 1000 bootstrap resamples of the data.

\section{Application to Super-K}
The Super-K detector~\cite{walter} is a 50 kton water Cherenkov detector located in the Kamioka  mine, which has been collecting data for more than 20 years, starting from April 1996 in four distinct phases. Super-K has detected neutrinos over six decades in an energy range from about 10 MeV~\cite{sksolar} to over a TeV~\cite{skshowering}.  The phase I of Super-K (sometimes referred to as  Super-K-I) lasted until July 2001, and has produced top quality experimental results in a diverse range of topics, including the solution of the solar neutrino  problem~\cite{sksolar01,sksolar02}, discovery of neutrino mass through atmospheric neutrino oscillations~\cite{sk98}, most stringent limits on relic supernova neutrinos~\cite{skrelic}, spin-dependent WIMP-proton couplings~\cite{skwimp}, neutrino magnetic moment~\cite{Liu}, proton decay~\cite{pd}, etc. 
In this paper we shall apply the generalized LS method to the first five  years of Super-K solar neutrino dataset, since it is publicly available. 
We first recapitulate  the methodology and  present the key results from the analysis done by the Super-K collaboration reported in Yoo et al~\cite{Yoo} (hereafter Y03), and then perform our own analysis of the Super-K data.

\subsection{Review of Super-K analysis}
Super-K has detected about 22,000 solar neutrino events between April 1996 and July 2001. 
Y03 did two periodicity searches with the Super-K data. The first analysis  was on the data divided into 5-day bins, and the second was on  the data  divided into 10-day bins. Both these analyses were done after correcting for the modulation due to  the eccentricity of the Earth's orbit around the Sun. Y03 applied the Lomb periodogram and then calculated the significance of the maximum peaks  using the method outlined in ~\cite{NR}. Their analysis does not take into account the errors in the flux per bin.  Other analyses of the same data have included the errors in fluxes. Moreover, Sturrock et al have also included the start and stop time of each bin and also the upper/lower error estimates~\cite{Sturrock07}. The asymmetric errors were also considered in the analysis done by Ranucci~\cite{Ranucci06}. These and other analyses of 
Super-K data have been done with many different methods, such as  wavelet-based analysis~\cite{Ranucci07}, Rayleigh Power analysis~\cite{Sturrock03}, maximum likelihood analysis~\cite{Sturrock05,Sturrock05a,Ranucci05,Ranucci06,Sturrock07}, etc. We briefly recap the analysis done by Y03 on the  dataset binned in 5-days.

Y03 scanned 36,000 frequencies between  0.0002/day  to 0.192/day. For this dataset, the maximum  power was found at 8.35 days with a confidence level of 63.1\% (or FAP of 36.9\%). Therefore, Y03 find no evidence for any statistically significant periods in both  these datasets, thus contradicting previous claims. 

\begin{figure}
\centering
\includegraphics[width=0.5\textwidth]{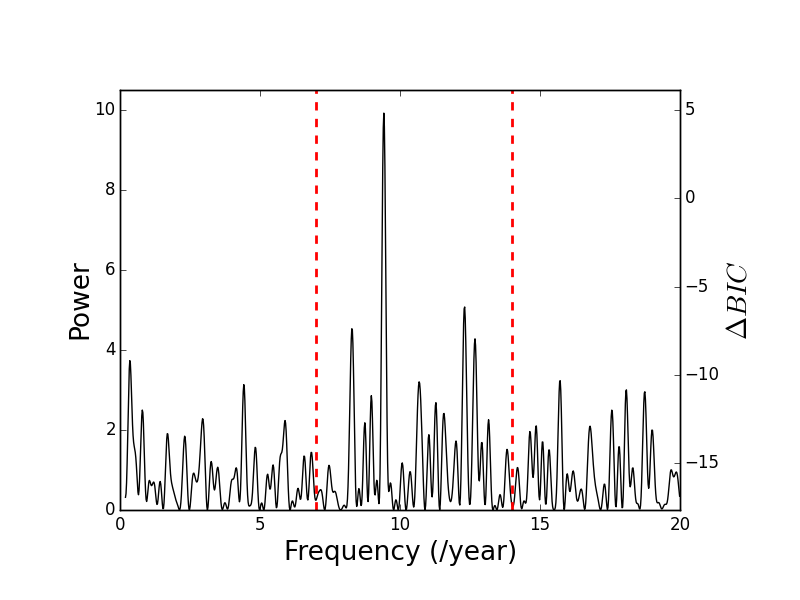}
\caption{The generalized LS periodogram applied to the Super-K dataset binned in 5-day intervals. The red dashed line shows the frequency range used to search for peaks. The right Y-axis on the bottom panel indicates the $\Delta$BIC, assuming the data is consistent with pure white noise, compared to the hypothesis that  the data consists of a sinusoid at  the given frequency (See Eq.~\ref{deltabic}). The maximum power is detected at a frequency of 9.42/year or period of 38.75 days corresponding to FAP of 1.5\% and $\Delta$BIC of 4.8. We note that significance level  depends on the range of frequencies scanned for peaks.}

\label{fig4}
\end{figure}

\subsection{Analysis of Super-K data}
We now apply the generalized LS periodogram as implemented in ~\cite{astroml} on the Super-K 5-day dataset. Similar  to Y03, we correct the measured flux for the eccentricity of the Earth's orbit around the Sun using the correction factors provided in Y03.  Unlike Y03, we consider the heteroscedastic errors for each bin, given by the average of the upper and lower errors.  The minimum frequency bin searched for is given by the reciprocal of the total duration of the dataset. Since we only want to search for any modulations caused by solar rotation, we restrict our searches
to a frequency range between 7/year (Period=50.76 days)  to 14/year (Period=26.11 days). This covers the variation in the synodic rotation rate from the solar equator to the poles~\cite{Beck2000}. We now present our results for the Super-K dataset, including our estimated significance for the first four peaks.

For this dataset, we scanned 358 frequencies between 7 and 14/year.  The LS periodogram is shown in Fig.~\ref{fig4}. The search band is delineated by the red dashed lines in this figure.
We find that the maximum  power occurs at a frequency of 9.43/year, corresponding to a period of 38.75 days. The FAP for this peak (from bootstrap resampling) is about  0.015 and  $\Delta$BIC is about 4.8.  The next set of peaks occur at 12.31/year, 8.24/year, and 12.66/year. However, $\Delta$BIC for all these peaks is less than zero, indicating that they are consistent with noise. Therefore, the only peak between 7-14/year which has $\Delta$BIC$>0$ is located at 9.43/year. The location, powers and statistical significance for the  first four peaks are tabulated in Table~\ref{table1}.

Sturrock et al  have analyzed the 5-day data in a number  of  papers over the past decade~\cite{Caldwell03,Sturrock03,Sturrock04,Sturrock05a,Sturrock08,Sturrock12} and have found that the maximum peak is same as in our analysis at 9.43/year. In 2005, they did multiple variants of the LS analysis of the 5-day day and found that the power has values of 11.67 and 11.24 (if floating offset is used) with FAPs of 0.7\%
and 1.9\%. They have also  evaluated the significance using generalized likelihood analysis, and  the resulting FAP  ranges between 0.02\%-1\%~\cite{Sturrock04,Sturrock07,Sturrock08}, with increasing significance as more information is incorporated.
From their latest analysis, they argue that the peak at 9.43/year is caused by  E-type $r$-mode oscillations due to Solar  rotation~\cite{Sturrock08,Sturrock12}. This peak corresponds to sidereal rotation rate of 10.43/year. They have also found other candidate peaks in the Super-K data at 8.3/year, 10.69/year, 11.57/year, 12.01, and 12.31/year corresponding to different spherical harmonic indices of $r$-mode oscillation~\cite{Sturrock08}. A peak at 9.42/year  with FAP of 1.8\% (12.1\% after including asymmetric errors) has also been detected from LS analysis by Ranucci of the same data~\cite{Ranucci05}.  However, we note that the FAP is a function of the frequency bandwidth used to search for periodicities. Both, 
Ranucci and Y03 searched a much larger frequency range for peaks and did not find the solar rotation related frequencies to be statistically significant. On the other hand, if the peak search is done in a smaller
region relevant to solar rotation and as more information gets incorporated, then the significance gets enhanced.  A tabular summary of powers along with significance from different LS analysis by various authors is included in Table~\ref{summarytable}.

 From our own analysis the only peak (detected by Sturrock et al), which is marginally significant from our analysis is the one at  9.43/year   corresponding to a period of 38.75 days. Using bootstrap resampling, we obtain a  FAP of  1.5\%.  For this peak,  the value of $\Delta$BIC is 4.8, which corresponds to positive evidence using the strength of evidence rules proposed in ~\cite{Shi}. In order for the peak to be considered real, $\Delta$BIC needs to exceed 10. 

 Therefore, the FAP which we get for this peak (using two different methods) is higher than that 
claimed by Sturrock et al and does not pass the 5$\sigma$ criterion used in particle physics experiments to claim detections.  We also checked the significance of this peak without eccentricity correction and it amounts to  $\Delta$BIC of 4.1, which again does not pass the 5$\sigma$ threshold. 
In the Appendix, we shall show that with this value of $\Delta$BIC, the amplitude of any sinusoidal modulation is less than 6\% of the mean solar neutrino flux. Nevertheless, if  this peak is real and corresponds to some new neutrino physics or some neglected physics in the Standard Solar Model, then the significance of this peak would be enhanced with accumulated analysis of the full  20  years of  Super-K solar neutrino data.

\begin{table}[t]
\begin{center}
\begin{small}
\begin{tabular}{ |l|c|c|c|c|}
\hline
Frequency & Period &    $\mathrm{P(\omega)}$ & FAP & $\Delta$BIC \\
(/year) & (days) & & & \\
\hline
9.43 & 38.75 & 10.53 &  0.015 & $4.8$ \\
12.31 & 9.34 & 5.0 & 0.17 & $-6.2$ \\
8.24 & 8.47 & 4.46 &  0.94 & $-7.4$ \\
12.66 & 5.84 & 4.23 & 0.98 & $-7.9$ \\
\hline
\end{tabular}
\caption{The significance  of first four peaks  in the Super-K 5-day solar neutrino data, calculated using two independent methods. The first two columns indicate the frequency and period of the peaks. The third column shows the  power.  The  fourth column contains the false alarm probability (FAP) for each frequency peak,  calculated using non-parametric bootstrap resampling~\cite{Suveges}. The last column shows the difference in BIC (Eq.~\ref{deltabic}) calculated in the same way is in the bottom panel of Fig.~\ref{fig4}. The most significant peak at period of 38.75 days coincides with that found by Sturrock. We note that FAP depends
on the search band used.}
\label{table1}
\end{small}
\end{center}
\end{table}

\begin{table}[t]
\begin{center}
\begin{small}
\begin{tabular}{ |l|c|c|c|}
\hline
Analysis & Peak Frequency (/yr) &    $\mathrm{P(\omega)}$ & FAP  \\ \hline
Yoo~\cite{Yoo} & 9.43 & 6.0 & 0.8 \\
Sturrock~\cite{Sturrock05a} & 9.43 & 11.67 & 0.007 \\
Sturrock~\cite{Sturrock05a}\footnote{Floating Offset method} & 9.43 & 11.24 & 0.019 \\
Ranucci~\cite{Ranucci05}  & 9.42 & 10.84 &  0.018 \\
Ranucci~\cite{Ranucci05}\footnote{Asymmetric errors}  & 9.42 & 9.22 &  0.12 \\
This Work (2016) & 9.43 & 10.53 & 0.015 \\
\hline
\end{tabular}
\caption{A summary of various Lomb-Scargle perodiogram analysis of solar neutrino datasets over the last
decade by various authors in the solar rotation frequency band (between 7-14 cycles/year). Note that for the first row, the peak frequency, power and FAP were extracted from Fig 6 of Y03~\cite{Yoo}.}
\label{summarytable}
\end{small}
\end{center}
\end{table}

\section{Conclusions}
Over the past two decades, there have been multiple claims of detections of  periodic modulations at different frequencies in the solar neutrino datasets from  Super-K, based on analysis by individual researchers of public data from its first five years. The general consensus based on analysis by Sturrock and collaborators is that these peaks are signatures of synodic rotation of the solar core~\cite{Sturrock08,Sturrock12}.
However, these results have not been confirmed by the analysis done by the Super-K collaboration  and so these claims remain controversial. Therefore, we do an independent search for any sinusoidal peaks in the Super-K dataset caused by solar rotation.

 In this paper, we use the generalized LS periodogram (proposed by Zechmeister and Kurster,~\cite{Kurster}), after accounting for the heteroscedastic errors,  on the publicly available solar neutrino datasets from Super-K (binned in 5-day intervals). This generalized periodogram (also called floating-offset periodogram in solar neutrino literature) is more sensitive than the regular periodogram. 
We then assess the significance of peaks at various frequencies  using two independent methods. The first is using non-parametric bootstrap resampling
and the second is by calculating the difference in Bayesian Information Criterion (BIC), assuming that the data is consistent with white noise and the data contains a peak at a given frequency. We note that, the generalized LS analysis is a special case of a likelihood analysis, where errors are symmetric and Gaussian. The BIC method, which we have used can also be applied to generalized likelihood based analysis to distinguish periodic signals from noise.

For Super-K, we performed our analysis on the first five years of data (publicly released in 2003),
divided into  5-day bins. We find a maximum peak at a frequency of 0.0258/day (period of 38.75 days). Its FAP is 1.5\% and $\Delta$BIC is about 4.8, which corresponds to positive evidence from the recommended strength of evidence rules for $\Delta$BIC~\cite{Shi}. Therefore, our estimated significance of this peak is marginal.  However, we note that previous searches, which did not find the peak to be significant scanned a very large range of frequencies. To look for peaks associated with solar rotation, it is important
to restrict the search to a narrow frequency range pertinent ot solar rotation, since the significance depends on the search band.

The same peak was detected at much higher significance by Sturrock et al  and the lowest FAP they obtain for this peak is 0.02\% after incorporating all available information and searching the right frequency range. Further analysis of the Super-K solar neutrino data (after combining data from all the four phases) is needed to check if this peak is real and if its significance increases with the accumulated data. All the remaining peaks found from our generalized LS periodogram analysis of the Super-K dataset are consistent with noise.

\begin{acknowledgements}
All the plots in this paper have been created using the {\tt astroML} Python library. We are grateful to the Super-K and SNO collaborations for making their solar neutrino datasets publicly available. Both the authors were part of the Super-K collaboration when Y03 was written and are grateful to our Super-K colleagues for many discussions related to this topic 13 years ago. We would like to thank the anonymous referee for many invaluable comments on the paper draft. Finally, we would  like to dedicate this paper to the memory of Danuta Kielczewska.
\end{acknowledgements}

\appendix

\section{Sensitivity Studies}

\begin{figure}
\centering
\includegraphics[width=0.5\textwidth]{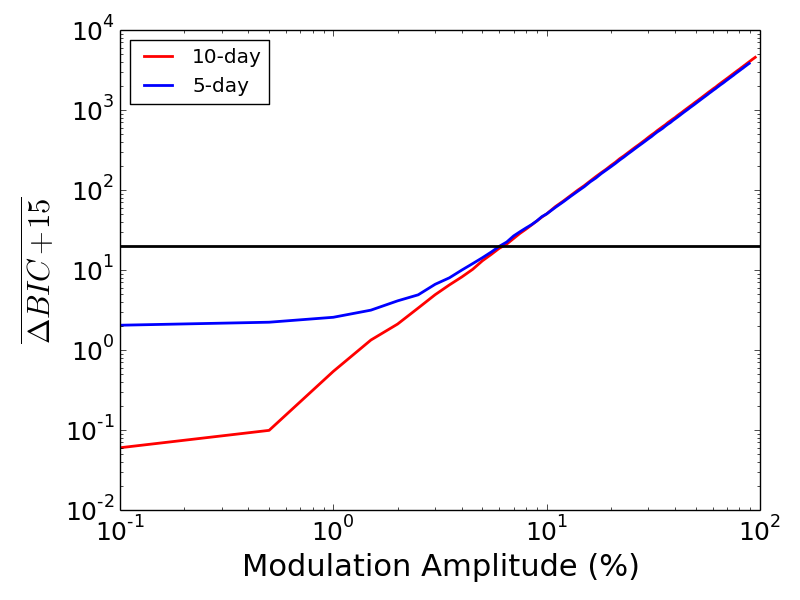}
\caption{Sensitivity to find periodicity at 0.0258/day (T=38.75 days) as a function of modulation amplitude using Monte-Carlo simulations of Super-K data using 5-day as well as 10-days bins. The Y-axis shows the average value of $\Delta$BIC (with an offset of 15 to enable using a log-log scale) for 1000 Monte-Carlo simulations for a simulated periodic signal at various modulation amplitudes. The horizontal line corresponds to $\Delta$BIC$=4.8$, which is the value we got from our analysis of the 5-day data. Therefore, a modulation amplitude of greater than 7\% would give a $\Delta$BIC value $>$10, corresponding to very strong evidence. From our observed value of $\Delta$BIC, the amplitude of the modulation signal is less than 6\% of the mean flux.}
\label{fig:simulation}
\end{figure}

We now study the feasibility of our  method to find periodicities as a function of modulation
amplitude (using $\Delta$BIC as the metric) for both the 5-day and 10-day datasets, by carrying out Monte-Carlo simulations.  Since we find a marginal peak at the $r$-mode frequency (0.0258/day), we focus our detection sensitivity studies only for this frequency. However, it is straightforward to extend this analysis to any other frequency, for which a peak can be detected from our LS analysis. Many similar Monte-Carlo studies of sensitivity analysis have been done in most analysis of  the public Super-K data by individual researchers (for eg.~\cite{Sturrock05,Ranucci06,Sturrock08} and references therein). We generate a time-series of the solar neutrino data consisting of a mean flux and a periodic signal with frequency of 0.0258/day. We varied the  modulation amplitude from 0 to 95\% of the mean flux. For each modulated signal, we carry out 1000 numerical experiments with the total signal in each bin generated from Gaussian fluctuations around the simulated flux with standard deviation   given by the flux error in each bin. We then apply the generalized LS
periodogram in exactly the same way  as for the real data. Since many previous simulation studies have used the false alarm probability as a metric to judge the significance, we only consider 
how $\Delta$BIC changes  with the modulation amplitude.  These results can be found in Fig.~\ref{fig:simulation}. As we can see, for modulation amplitudes less than $\sim$ 6\%, the 5-day binned analysis would be more sensitive to any observed modulations compared to the 10-day analysis. For modulation amplitudes larger than this, both analysis are equally sensitive.
 From this figure, a modulation amplitude between 6.5\%-7\% corresponds to a $\Delta$BIC value indicating strong evidence, and a modulation amplitude $>7$\% corresponds to very strong evidence. From the observed value of $\Delta$BIC of 4.8  (for the 5-day Super-K dataset), we conclude that the modulation amplitude of any periodic signal is less than 6\% of the mean solar neutrino flux. 

\section{Periodogram analysis of SNO data}

The Sudbury Neutrino Observatory (SNO) was another solar neutrino experiment, which collected data from 1999 to 2006 using heavy water (\Dtwoo) as the target. It  provided the definitive solution to the long standing solar neutrino puzzle~\cite{sno16} and measured the total flux of $^8B$ neutrinos to a much better accuracy than those predicted by the Standard Solar Model. SNO consisted of three phases depending on how the total neutral current reaction rate was measured.  More details on the key physics results from the SNO experiment can be found in Ref.~\cite{sno16}. In this work, we report analysis of the publicly available SNO data from both the pure \Dtwoo phase~\cite{d2o} (Phase-I) as well as the salt~\cite{salt} phase (Phase-II) in which two tonnes of NaCl were added to increase the neutral current detection efficiency.
 The SNO collaboration did two periodicity searches with their data. One was a 
low-frequency search~\cite{SNO} followed by a high-frequency search~\cite{SNO09}.
We first briefly recap the low-frequency analysis done by the SNO collaboration~\cite{SNO}, and then present our analysis of the same data.

We now do the same LS analysis on the SNO data.
The SNO collaboration did a separate analysis of the \Dtwoo as well as the salt datasets. The \Dtwoo dataset spans about 572 days in which 2924 neutrino events were detected. The salt dataset spans a livetime of about 763 days in which 4722 neutrino events were detected.
The periodicity analysis was done after splitting both the datasets into 1-day bins. Two types of searches were done. The first  was an unbinned
maximum likelihood analysis. The maximum significance (with this method) was found at a period of 3.5 days for the \Dtwoo dataset and at 1.03 days for the salt dataset. However, the FAPs for these peaks are 35\% and 72\% respectively, thereby indicating that these  are not statistically significant. The second method involved  the   LS periodogram analysis of  both the datasets, after accounting for the heteroscedastic errors in each data point. A total of  7300 frequencies were scanned from 10 years down to the Nyquist limit of two days. The largest peak was found for the \Dtwoo dataset at 2.45 days and for the salt dataset at a period of 2.33 days. However, the FAPs for these peaks are 46\% and 65\% respectively, which is consistent with noise. More details on these results can be found in Aharmim et al~\cite{SNO}. Subsequently, a high-frequency search of the same dataset was done using Rayleigh Power analysis where all frequencies from 1/day to 144/day were scanned for periodicities~\cite{SNO09}. But no, significant peaks were detected in this high frequency search.
Therefore, the analysis done by the SNO collaboration found no evidence for any statistically significant peaks, including those found by Sturrock and collaborators in the SK data.

Both the salt and \Dtwoo data are provided with one-day binning. Similar to the SNO collaboration, we calculated the fractional livetime
of each day between the first and last  neutrino detection. The number of neutrinos in each bin was normalized by this livetime to get the total rate per bin. We used 68\% Poissonian error bars for each bin~\cite{Gehrels86}. Note that in the analysis done by the SNO collaboration~\cite{SNO}, all bins with low statistics were merged with the neighboring bins. Here, we do not merge bins with low counts.
 For both the datasets, we scanned a total of 5740 frequencies between 0.00174/day and 
1/day. Similar to our Super-K analysis, we choose the maximum frequency to be twice the Nyquist value. 
This frequency range would be sensitive to any periodicities detected in Super-K data. However, with 1-day binning, our LS analysis is not sensitive to the high-frequency search done in ~\cite{SNO09}.
We evaluate the statistical significance using non-parametric bootstrap resamples of the data, and by calculating $\Delta$BIC from Eq.~\ref{deltabic}. For the \Dtwoo data, the maximum peak occurs at a frequency of 0.257/day. The value of $\Delta$BIC is 3.9 and the value of FAP is 22\%. For the salt data, the maximum peak occurs at a frequency of 0.0427/day corresponding to a FAP of 43\% and $\Delta$BIC of 1.03. Therefore, the peaks  in the LS periodogram are not significant.

Ghosh et al~\cite{Ghosh} have analyzed the same data from SNO using Date Compensated Discrete Fourier Transform technique  (which is similar to the LS periodogram) and   have detected  peaks with more than 95\% confidence level (or FAP $<5\%$) at 5.36, 14.99, 23.06, 40.31, 65.63, 115.03, 131.27, 152.67, 188.74, 294.49, and 412.79 days in the \Dtwoo phase, and at 37.84, 43.89, 59.17, 67.48, 72.47, 78.52, 86.54, 95.91, 109.83, and 317.01 days in the salt phase. We calculate the statistical significance of these frequencies using both the methods. The results are shown in Table~\ref{table2}. For all these frequencies, our estimated FAPs are close to  one and all the $\Delta$BICs are less than zero. Therefore, none of the reported peaks claimed by Ghosh et al~\cite{Ghosh} in the SNO data are statistically significant. We also note that the regular LS periodogram analysis
of the SNO data has been independently done in ~\cite{Ranucci06,Ranucci07}, and they also fail to find any statistically significant peaks. Sturrock et al also searched the SNO \Dtwoo data to check for modulations at a frequency of 9.43/year (found in the 5-day Super-K dataset), but could not find this peak~\cite{Sturrocksno}.

\begin{figure}
\centering
\includegraphics[width=0.5\textwidth]{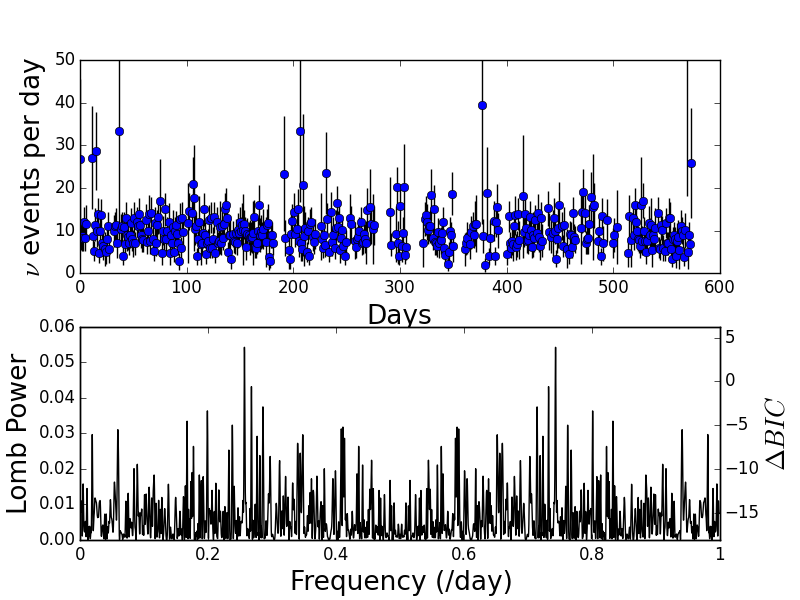}
\caption{The generalized LS periodogram applied to the SNO data~\cite{SNO} from the salt phase. The top panel shows the rate of solar neutrino events with Poissonian error bars. The bottom panel shows the generalized  LS periodogram applied to this data as a function of frequency along with $\Delta$BIC (between pure noise and the peak at a given frequency) values on the right axis. The maximum power is detected at a frequency of 0.257/day (or period  of 3.89 days) corresponding to FAP of 22\% and $\Delta$BIC of 3.9. Note that the periodorgams have been normalized according to Ref.~\cite{astromlbook}.}
\label{fig5}
\end{figure}

\begin{figure}
\centering
\includegraphics[width=0.5\textwidth]{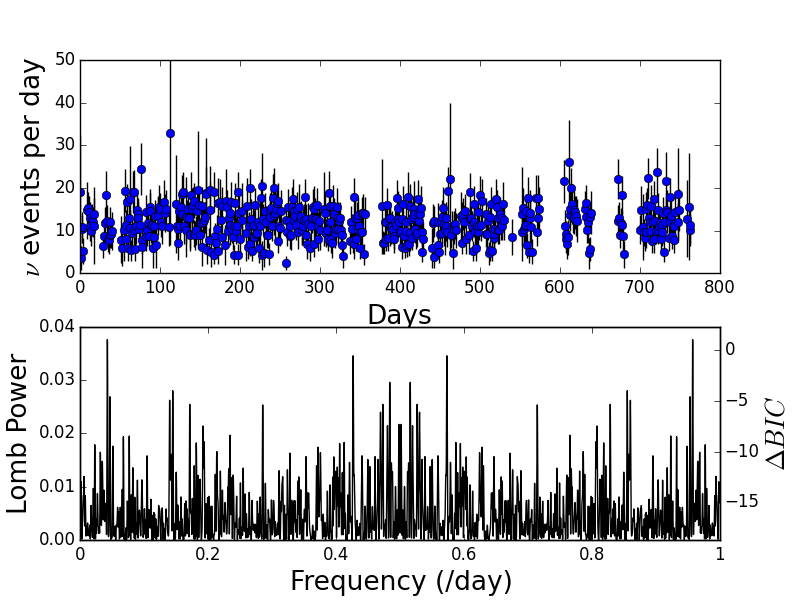}
\caption{The generalized LS periodogram applied to the SNO data from the salt phase. All the axes in both the panels are the same as in Fig.~\ref{fig5}. The maximum power is detected at a frequency of 0.043/day (or period  of 23.25 days) corresponding to FAP of 43\% and $\Delta$BIC of 1.03.}
\label{fig6}
\end{figure}

\begin{table}
\begin{center}
\begin{small}
\begin{tabular}{ |l|c|c|c|c|}
\hline
Dataset & Frequency(/day) &  Period (days) & FAP & $\Delta$BIC \\
\hline
\Dtwoo & 0.257 & 3.89 & 0.22 & $3.9$ \\
& 0.186 & 5.36 & 1 & $-10.7$ \\
& 0.0667 &  14.99 &  1  & $-1.6$ \\ 
&  0.0433 &  23.06 &  1  & $-1.6$ \\ 
& 0.0248 &  40.31 & 1  & $-1.6$ \\
&  0.015 &  65.63 & 1  & $-1.6$ \\
& 0.0086 & 115.03 & 1  & $-1.6$ \\
& 0.0076 & 131.27 & 1  & $-1.6$ \\
& 0.0053 & 188.74 & 1  & $-1.6$ \\
& 0.0033 & 294.49 & 1  & $-1.6$ \\
& 0.0024 &  412.79 & 1  & $-1.6$ \\ \hline
Salt & 0.0427 & 10.04 & 0.43 & $1.03$ \\
& 0.026 & 37.84 & 1 & $-16.4$ \\
& 0.023 & 43.89 & 1 & $-10$ \\
& 0.017 & 59.17 & 1 & $-15.7$ \\
& 0.0148 & 67.48 & 1 & $-17.4$ \\
& 0.0137 & 72.47 & 1 & $-18.2$ \\
& 0.0127 & 78.52 & 1 & $-18$ \\
& 0.012 & 86.54 & 1 &  $-17.3$ \\
& 0.010 & 95.91 & 1 &  $-17.8$ \\
& 0.009 & 109.89 & 1 & $-18$ \\
& 0.003 & 317.01 & 1 & $-18.2$ \\
\hline
\end{tabular}
\caption{The significance  of various peaks  in the SNO datasets from both the salt and \Dtwoo phase using two independent methods calculated in the same way as in  Table~\ref{table1}. The first row for each of the datasets indicates the peak frequency from our generalized LS analysis along with its significances. The rest of the table contains the significance  at various  frequencies for which statistically significant peaks have been reported in Table 1 of Ref.~\cite{Ghosh}. Therefore, our analysis shows that all the reported peaks in Ref.~\cite{Ghosh} are consistent with noise.}

\label{table2}
\end{small}
\end{center}
\end{table}
\end{document}